# Organization of the Bacterial Light-Harvesting Apparatus Rationalized by Exciton Transport Optimization

Elad Harel

Northwestern University, Department of Chemistry, 2145 Sheridan Road, Evanston, IL 60208

Correspondence and requests for materials should be addressed to

Elad Harel

2145 Sheridan Road

Technological Institute K344

Evanston, IL 60202

847-467-7580

elharel@northwestern.edu

## Abstract

Photosynthesis, the process by which energy from sunlight drives cellular metabolism, relies on a unique organization of light-harvesting and reaction center complexes. Recently, the organization of light-harvesting LH2 complexes and dimeric reaction center-light harvesting I-PufX (RC-LH1-PufX) core complexes in membranes of purple non-sulfur bacteria was revealed by atomic force microscopy (AFM)[1]. Here, we report that the structure of LH2 and its organization within the membrane can be largely rationalized by a simple physical model that relies primarily on exciton transfer optimization. The process through which the light-harvesting complexes transfer excitation energy has been recognized to incorporate both coherent and incoherent processes mediated by the surrounding protein environment. Using the Haken-Strobl model, we show that the organization of the complexes in the membrane can be almost entirely explained by simple electrostatic considerations and that quantum effects act primarily to enforce robustness with respect to spatial disorder between complexes. The implications of such an arrangement are discussed in the context of biomimetic photosynthetic analogs capable of transferring energy efficiently across tens to hundreds of nanometers.

Introduction

The first step in photosynthesis is the absorption of light by the light-harvesting (LH) apparatus[2]. Transfer of energy from the LH to the reaction center (RC) leads to a stabilized charge-separated state across the membrane, which drives chemical transduction. The high symmetry of the pigment-protein complexes that compose the LH apparatus of bacterial photosynthesis as revealed by high-resolution x-ray crystallography[3] has motivated extensive theoretical and experimental investigations in an attempt to understand both the biological[4] and physical[5] significance of the structure of these complexes and their organization within the membrane. The photosynthetic bacterium *Rhodobacter sphaeroides* serves as a model system to understand the functional role of this organization. The optical spectroscopy[6-10] of individual pigment-protein complexes in these bacteria has revealed that both quantum and classical transport phenomena may play a role in the remarkable efficiency of energy transfer in these systems. Theoretical work on these complexes at a high level of detail[9,11,12] has attempted to understand the influence of quantum effects on both intra- and intermolecular transfer within and between complexes.

In this report, we examine the implications of assuming that the structure of these complexes is optimized for efficient photo-capture and transport. We use a simplified model that neglects quantum coherence and treats the coupling between pigments by simple dipole-dipole interactions only. By examining large structures of chromophores arranged in different topologies, we explain the most salient features of the LH apparatus: nonomeric symmetry of LH2, ring-like structure, inter-complex distance, inter-complex topology, and inter-complex transfer time. Critically, our model takes into account only the most basic features of the complexes, allowing us to unambiguously identify the core principles responsible for its primary function – light harvesting and energy transfer. Furthermore, we demonstrate that quantum effects do not necessarily increase efficiency of transfer, but, rather, make transfer more robust to disorder – a critical feature of the LH apparatus necessary to operate efficiently in the 'hot and wet' environment of the organism. Finally, we briefly discuss approaches to mimic the high quantum efficiency and large exciton diffusion lengths that may significantly increase performance of current artificial photosynthetic constructs.

## Theory

We consider a simple model in which an exciton system is coupled to a bath, ground state, and trap. This is essentially the Haken-Strobl model. We treat the problem by formalism nearly identical to the one used by Cho and Silbey in a recent publication[13] in which they investigated simple linear chains and simple, nonlinear topological arrangements of loops with three and four sites. The master equation is given by

$$\dot{\rho}(t) = -[\mathcal{L}_{sys} + \mathcal{L}_{dissip} + \mathcal{L}_{decay} + \mathcal{L}_{trap}]\rho(t) \qquad (1)$$

where $\rho$ is the density matrix and $\mathcal{L}$ are Liouville operators. The system consists of N dipoles at positions, $r_i$, including one trap state, $i = tr$, coupled by the dipole-dipole interaction, $V_{ij}$. For the sake of simplicity the pigments will be assumed to lie in a plane and their dipoles will all lie parallel to one another, normal to the plane. While in LH2 this is not the case (dipoles are in a circular arrangement), the consequences for energy transfer and LH organization are negligible. Furthermore, while the membrane structure of Rh. sphaeroides is a vesicle and not a plane, this effect is also negligible in the context of our analysis. A simpler arrangement of dipoles allows us to more easily transition to a discussion of possible designs for biomimetic systems that use repeating units of pigments.

For the remainder of this work, the decay part of the master equation will be ignored since $k_d \ll k_t$ for the systems examined here, where $k_t$ represents the average trap rate and $k_d$ represents the relaxation rate to the ground state. $\mathcal{L}_{trap}$ represents irreversible decay to a trapping state, e.g. a charge separated state in the photosynthetic reaction center. It is expressed as $[\mathcal{L}_{trap}]_{nm} = (\delta_{tr,n} + \delta_{tr,m})k_t/2$. The factor of two comes from the fact that the theoretical limit for relaxation in a population is twice that of a coherence according to Redfield theory[14]. Finally, the dissipation is modeled as $[\mathcal{L}_{dissip}]_{nm} = \Gamma^*_{nm}$. To simplify matters more, we make the stationary approximation in which the coherence terms are time-independent. More accurately, we assume that coherences dephase much more rapidly than the time scale of population relaxation, an approximation validated by recent experimental results on isolated LH2 complexes using coherent two-dimensional optical spectroscopy[15]. With this assumption, the master equation can be written as

$$-i \sum_{j\neq n,m} (H_{nj}\rho_{jm} - H_{jm}\rho_{nj}) - \left\{(1-\delta_{nm})\Gamma^*_{nm} + \frac{1}{2}(k_t\delta_{n,tr} + k_t\delta_{m,tr}) + i\Delta_{nm}\right\}\rho_{nm} = \dot{\rho}_{nm}\delta_{nm} + iH_{nm}(\rho_m - \rho_n) \quad (2)$$

where $\Delta_{nm} = \epsilon_n - \epsilon_m$ is the detuning and $\epsilon_i$ represents the energy of the $i^{th}$ site. $H$ is the system Hamiltonian in the site basis. This equation can be conveniently expressed in block matrix form as follows.

$$\begin{pmatrix} 0 & B \\ B^T & K \end{pmatrix} \begin{pmatrix} \rho_P \\ \rho_C \end{pmatrix} = \begin{pmatrix} \dot{\rho}_P \\ 0 \end{pmatrix} \tag{3}$$

where $\rho_P = (\rho_1, \rho_2, \dots, \rho_{tr-1}, \rho_{tr+1}, \dots, \rho_N)$ and $\rho_i \equiv \rho_{ii}$ describes the population at each site, except for the trap state which is dealt with separately, and $\rho_C = (\rho_{12}, \rho_{13}, \dots, \rho_{1N}, \rho_{21}, \rho_{23}, \dots, \rho_{2N}, \dots, \rho_{N1}, \rho_{N2}, \dots, \rho_{N-1,N})$ describes the coherence terms. Construction of the *(N - 1) x (N*$^2$ *- N)* matrix, *B*, and the *(N*$^2$ *- N) x (N*$^2$ *– N)* matrix, *K*, are described in the appendix for sake of continuity. From (3), one gets that $B\rho_C = \dot{\rho}_P$ and $B^T\rho_P + K\rho_C = 0$. Solving these two equations gives

$$\dot{\rho}_P = -BK^{-1}B^T\rho_P \tag{4}$$

Integrating both sides and assuming that at very long times the population of the system has all decayed to the trap state, results in

$$\rho_P(0) = BK^{-1}B^T\hat{\tau} \tag{5}$$

where $\hat{\tau} = (\tau_1, \tau_2, \dots, \tau_{tr-1}, \tau_{tr+1}, \dots, \tau_N)$ and $\tau_n \equiv \int_0^\infty \rho_n(t)$ is the average residence time of excitation at site *n*. At the trap state, the average residence time is simply given by $\tau_{tr} = 1/k_t$ (see Appendix). Solving for the average residence time at each site requires inversion of an *(N*$^2$*- N) x (N*$^2$ *– N)* matrix, followed by inversion of an *(N - 1) x (N - 1)* matrix. Integrating out the time dynamics simplifies the analysis and computational demands considerably; For instance, for *N* = 100 sites the average residence time can be calculated in under 20 seconds using a quad-core workstation. The quantum efficiency of transport to the trap is then given by

$$q \approx \frac{1}{1 + k_d\langle t\rangle} \tag{6}$$

where $\langle t\rangle \equiv \sum_{i=1}^{N} \tau_i$ is the average trapping time.

Classical transport may also be calculated using this basic formalism. For any N-site system, the classical rate equations are given by

$$\dot{\rho}_n = -\sum_{i\neq n}^{N} k_{n,i}(\rho_n - \rho_i) - \delta_{n,tr}\rho_{tr} \tag{7}$$

where the classical rate constant is given by

$$k_{ij} = \frac{2\Gamma_{ij}}{\left(\Gamma_{ij}^{*} + \frac{k_t}{2}(\delta_{i,tr} + \delta_{j,tr})\right)^2 + {\Delta_{ij}}^2}\left|V_{ij}\right|^2 \tag{8}$$

In this context, $\Gamma_{ij}$, is the line width of the transition from site *i* to site $j$. Again, one can solve this by simple matrix inversion to find the classical average residence time at each site. Unlike the quantum case, which scales as the square of the system size, the classical rate equations scale linearly with the size of the system, therefore requiring an *N x N* matrix inversion.

## Results and Discussion

The organization of this section is as follows. First, we show why ~9 fold symmetry is optimal for a ring-like structure of a given size (5 nm diameter in this case). Second, we show that nearly degenerate site energies lead to an optimal trapping time – strongly suggesting that utilizing identical chromophore units is advantageous. Next, we show that ring-like structures are advantageous in terms of structural disorder relative to linear chains or other topologies. This result is then related to the optimal organization of complexes in a staggered or packed, rather than linear arrangement. We also discuss the whether the organization of these complexes is highly tuned. Finally, we contrast purely classical and mixed quantum/classical transport.

The organization of the LH apparatus on spherical vesicles[16] is shown in Figure 1. LH2 consists of two ring structures, named the B800 and B850 ring for their respective absorption bands in the near infrared (NIR). The diameter of a single, nonomeric LH2 complex is about 8 nm[17]. Ignoring the protein scaffold, the largest intra-ring distance for either ring is about 5 – 5.5 nm. Using a genetic algorithm, we minimized the quantum average trapping time as a function of the spectral detuning between sites, $\Delta_{nm} = \epsilon_n - \epsilon_m$, the dephasing matrix, $\hat{\Gamma}^*$, and the trap rate, $k_t$, for three staggered, 5 nm rings containing $N$ = 4 -14 elements each. The optimal energies and dephasing terms between coupled rings are shown in Figure 2. The coupling strength between site dipoles as given by the magnitude of the off-diagonal matrix elements is shown schematically by lines connecting sites. In each case, the donor and trap (i.e. acceptor) states were chosen so at to keep the transfer distance approximately constant. Besides the number of elements per ring, the topology remained the same for each case. A plot of the optimal average transfer time against the number of elements per ring shows that the minimum transfer time is reached around 8-9 elements per ring, with a tapering off at higher number of elements. From a biological perspective, utilizing the smallest number of elements that achieves the optimal performance minimizes the biochemical costs associated with creation of unproductive chromophores. It is also important to note that as the number of elements per ring reaches $N > 6$, each ring begins to act like a single entity. This stems from the fact that when the coupling between rings is stronger than the detuning between sites, the excitation becomes delocalized across the rings. The orientation of the dipoles within each complex, as long as it preserves circular symmetry, is therefore not terribly important as this acts to modulate the intra-chromophore coupling but does not change the dynamics or have noticeable influence on the organization of the LH apparatus. The delocalization feature is a critical element of the design strategy used by the LH apparatus as illustrated in the next section.

We then explored the question of topology – whether a ring structure for LH2 was really compared to some other unexplored geometry. Consider the five rings shown in top panel of figure 3. Exciton diffusion through this large complex takes about 80.5 ps after optimization of the detuning, dephasing, and trap rate. The optimal trapping time is reached when the site energies are almost completely degenerate, while the optimal dephasing rate appears to vary from about 0 – 3 $cm^{-1}$ (maps not shown). Now consider the same length structure, but this time with single sites instead of ring structures. Again, the optimal trapping rate is reached when the site energies are nearly degenerate and with very low values of the dephasing rate. This is in agreement with classical rate equations for linear chains, since the classical and quantum limits in this geometry are nearly identical[18]. The average transfer time for the linear chain is only 21.5 ps, about one quarter that of the ring structures. This suggests that linear chains are more efficient, which intuitively makes sense since the excitation spends less time on sites that do not directly participate in transfer. Now consider what happens when we incorporate spatial disorder, $\Delta r$, in the system. At 0.5 nm disorder ($\Delta x_{max} = \Delta y_{max} = 0.5$ nm) in the plane, the transfer time in the linear chain increases by nearly five-fold to about 102 ps. For 1 nm disorder, it dramatically increases to 1.2 ns and with 2 nm disorder it reaches 6.2 ns. By stark contrast, in the ring chain, 0.5 nm disorder increases the transfer time by about 50% to 113 ps, while 1 nm disorder increases it to only 136 ps. 2 nm disorder raises this further only to 164 ps.

We then repeated these runs multiple times (5 time for the ring and 20 times for the linear chains) to obtain better statistics and understand the general trends. Figure 4 shows the dependence of the degree of average trapping time on the spatial disorder. For the linear chain of sites, the dependence is almost perfectly matched with a cubic dependence on disorder. This may be explained by the cubic dependence of the dipolar interactions with distance between chromophores. The rate goes as the sixth power, but 'diffusion' in one-dimension goes as the square root of time, giving an effective cubic dependence. That is, each instance of random spatial disorder starting from an initial linear distribution of sites, gives an average distance between sites that scales in a manner similar to a quasi-one dimensional diffusing particle (i.e. sites can move in either direction, but initiate with a linear configuration). Also, notice that for the linear chain the error increases almost linearly with spatial disorder – another feature that may be explained by analogy to one-dimensional diffusion. In contrast, the dependence of the trapping time with spatial disorder is approximately linear in the regime examined for the ring case. These results clearly indicate that ring structures are more robust to disorder. Effectively, each ring acts like a single chromophore such that a given amount of disorder appears smaller to a large chromophore than to a smaller one. Therefore, it is advantageous to use large chromophores for the sake of robustness, at least up to a point. If a ring gets too large then the exciton diffusion length begins to suffer in proportion to the ring diameter. One may ask why not adopt some grid-like structure, or a ring that is partially filled in with chromophores. As shown in Figure 5, such an arrangement wastes efficiency as excitation is spent on sites that do not participate in direct transfer processes. Or to put it another way, crossing a ring scales as the diameter of the ring, while using the interior scales with the square. Therefore, a ring-like structure is optimal given the high symmetry needed to transfer energy in any given direction, combined with the inherent robustness utilizing larger effective units.

The linear ring chain arrangement brings up another question of what is the best way to arrange the rings themselves with respect to one another. As can be appreciated from the analysis above, disorder inevitably creates regions in which the electrostatic coupling between two sites is too weak for efficient transfer, i.e. the rate limiting step comes about from “breaks” in the linear chain. Since each ring is itself acting like a single chromophore, the same argument can be made for a linear chain of rings as shown in Figure 6. However, if the rings are staggered, then disorder is less likely to produce such breaks. Or to put it more succinctly, a ring of rings is optimal for the same reason that the ring was optimal to begin with. However, a ring of rings does not provide a clear means by which to integrate a single reaction center, so a staggered or packed arrangement is a suitable compromise. Incorporating random disorder to the ring centers and less than 0.1 nm disorder to the elements of the rings themselves, demonstrates the robustness of the average trapping time in this organization (see Figure 3 and accompanying discussion). Disorder in the elements of the ring is small in large part because of the relatively stiff protein scaffold, which holds the chromophores in a precise arrangement due to electrostatic interactions with nearby amino acid residues. The electrostatic interactions between the rings also act to increase the bandwidth of absorption beyond that possible by individual ring. The energy spectra resulting from diagonalization of the system Hamiltonian for the packed and linear arrangements are also shown in figure 6 and qualitatively reproduce the energy spectrum observed in the B800 band of single LH2 complexes[5,19].

The approach we have used thus far is to optimize the minimal trapping time as a function of a relatively large parameter space involving the energies at each site and ~$N^2$ dephasing terms between pairs of sites as well as the trapping rate. This brings up the question of whether natural photosynthetic systems have somehow tuned all of these parameters through an evolutionary process. Intuitively, it may be hard to believe that evolution could optimize so many parameters while still remaining functional. To explore this question in depth, we ran an optimization procedure for the 3-ring structure shown in figure 7. Repeated runs to insure proper convergence of our optimization scheme gave a value of 63 ± 2 ps average trapping time from donor at site 6 to acceptor at site 10. We then ran the same model (no optimization), but now fixed all the energies to the same value and set all the dephasing terms to be equal: $\boldsymbol{\Delta_{ij}} = \mathbf{0}$ and $\boldsymbol{\Gamma_{ij}}^{*} = \boldsymbol{\Gamma}^{*}$. As shown in Figure 7c, the optimal trapping time is reached at $\boldsymbol{\Gamma}^{*} = \mathbf{10}$ $cm^{-1}$ and $\boldsymbol{k_t} = \mathbf{110}$ $cm^{-1}$, giving a trapping time of 71 ps, which is only 10% higher than the optimal value. This suggests that the system is not highly tuned and that only a two, rather than ~$N^2$ parameters, are necessary to achieve near-optimal transport. These results indicates the a robust, artificial system may be fabricated without molecular-level control of the site environment that would otherwise need tuning through specific interactions with a tailored spectral density.

Our results are also consistent with the results of other works[20-25], in which finite dephasing assists the transport since it eliminates stationary states that do not couple to the trap state. It is also interesting to point out that the optimal trapping time configuration predicts about 4-5 picoseconds per site. Given the delocalization of the exciton, this number is consistent with photon echo measurements[26] that show a ~5 ps decay of B850 band in intact membranes of *Rh. Sphaeroides*. Without adequate spatial or spectral resolution such spectroscopic measurements cannot straightforwardly distinguish decay at one site from decay of a delocalized state. These results demonstrate that this very simple model captures most of the salient features of the spectral and dynamical properties of the system without invoking quantum coherence, details of the high-level structure of the complex, or identification of the spectral density that couple the pigments to the bath.

Next, we considered the role of quantum mechanics in the exciton energy transfer compared to purely classical transport mechanisms for the 3-ring structure shown in figure 8. Interestingly, the quantum model shows a slower optimal transfer time by about a factor of three when compared to the optimal solution found using the classical model (red and blue arrows in the inset plot). At first glance, it appears then that quantum mechanics does not aid in transfer. However, when considering the issue of robustness a more complicated picture emerges. Again, consider the case in which the site energies are degenerate and the dephasing rate (i.e. line width for the classical case) between all sites is identical (Figure 8A). The optimal transfer time in the classical picture is at near-zero values of the line width and at a trap rate of ~125 $cm^{-1}$. Again, this picture is consistent from consideration of the denominator of equation (8). In the quantum case, the optical transfer time is reached at about the same trap rate, but at a finite dephasing rate of about 10 $cm^{-1}$. As before, tuning only these two parameters produces nearly the same transfer times as the full optimization. Now, consider the case when there is energetic disorder in the system (Figure 8B). Here, we have added a random detuning up to a maximum of 100 $cm^{-1}$ at each site. In the quantum case, the trapping time map is remarkably similar in the presence of this type of disorder. However, for the classical case, the line width shifts from near-zero in the degenerate case to about 33 $cm^{-1}$ in the disordered case. Changes in site energies cause dramatic changes in the optimal line width for purely classical transport, while for quantum transport the system is significantly more robust. From a biological perspective this strategy is advantageous because the dephasing rate / line width is set by the details of the protein environment. Quantum mechanics acts to overcome changes in the energy landscape of the system that appear as kinetic traps in classical transport phenomenon, primarily through delocalization and through mechanisms that allow transport even between sites that are not directly coupled. These non-local kinetic networks were recently explored by Cho and Silbey for simple, linear chains and three- and four-level non-linear topologies[13].

Of course, the LH2 complex is composed of two rings, not one. Therefore, it is worth commenting on the possible origins and benefits of incorporating two rings. Purple non-sulfur bacteria do not utilize a chlorosome antennae complex as do green sulfur bacteria. The chlorosome consists of many tens to hundreds of thousands of photosynthetic pigments, but is devoid of a protein matrix. The purpose of the chlorosome is simply to act as a broadband and large surface area antenna[2] to capture solar flux and direct it to the reaction center through an intermediate pigment-protein complex called the Fenna-Mathews-Olson (FMO) complex. The chlorosome funnels energy to populate one specific state, delocalized about ~2 sites, in the FMO complex. In the absence of the chlorosome, the LH apparatus of purple bacteria utilizes localized and broadband antennae – namely, the B800 rings. These antennae may simply serve the same purpose as the chlorosome, but with a distributed rather than localized architecture. Energy absorbed by the B800 ring transfers to the B850 ring in about 1 ps and is then directed towards a single acceptor site. Due to the strong intra-chromophore coupling in the B850 ring combined with static and dynamic disorder in the complex energy rapidly (< 300 fs) localize on one specific site. From this point forward, transfer may be described by the model used in this work. The origin of whether or not to utilize a chlorosome to begin with, however, is most likely driven by other biological considerations that have little or nothing to do with energy transfer.

Finally, we discuss the implications for energy transfer in biomimetic systems based on these results. The fact that fine-tuning is not a necessary feature for near-optimal transfer is promising; using repeating units with nearly degenerate site energies and nearly equal dephasing rates is significantly easier than controlling the system-bath interactions at each site. Depending on the synthetic method used to piece together individual chromosomes into large units such as rings, the relative orientation of the transition dipoles may vary. Creating rings with different properties may be achieved by capitalizing on relative dipole orientations or by using of different constituent dipoles, number of elements, and local scaffold environment. Fortunately, the exact details of the chromophores are less important than the delocalized states that they form. Therefore, many candidates may be utilized as chromophores and trap states (i.e. charge separated species). Controlling the system-bath interactions – the dephasing rates in our model – may be achieved by using a polymer host with varying degree of cross-linking and, therefore, stiffness. Such constructs are yet to be explored in this context.

## Conclusion

We have shown that the most salient features of the organization of the LH apparatus in model bacterium can be explained with the Haken-Strobl model without consideration of the high-level details of LH2 and LH1-RC complex. Integrating out the dynamics and considering only the mean residence time at each site simplifies the analysis considerably, allowing us to focus on the basic physics. By utilizing an optimization scheme, we have shown that rings are ideal structures by which to transfer energy efficiently for several reasons: 1) the high symmetry allows transfer in any direction, which is critical when the location of the RC (i.e. trap) is not known a priori, 2) the ring essentially acts to make the chromophore larger, increasing robustness to disorder among complexes, 3) filling in the interior of the structure acts to waste precious biochemical resources by spending time at sites that do not participate in efficient transfer across space; hence, a void inside the structure is ideal. Taken together, these features uniquely define a ring. We have also shown that a composite of rings in a staggered or packed configuration is more robust to spatial disorder as it avoids the likelihood of breaks that are the rate-limiting step in the transfer process. We demonstrate that while our optimization schemes may give the lowest transfer time, the system need not be fine-tuned to reach a comparable transfer time. This observation makes designing artificial photosynthetic complexes much more promising than if fine-tuning was necessary to achieve high quantum efficiency. Nature's strategy to utilize repeating units of identical chromophores and place them in a single protein host is advantageous from a biological as well as physical perspective. Finally, we showed that quantum mechanical effects, at least based on this simplified model in which quantum coherence or pathway interference is effectively ignored, does not necessarily aid in transfer efficiency. Rather, it increases the robustness of the system to a fixed value of the dephasing and trap rates. Natural systems cannot dynamically tune these parameters in response to changes in energetic disorder since they are inherent to the pigment-protein structure. From a biological point-of-view, robustness is oftentimes more important than overall efficiency, especially when the decay to the ground state is so much slower than the time scale of transfer. Remarkably, the crude model employed here, which neglects the high-order structure, predicts the correct decay times observed in nonlinear spectroscopic experiments between LH2 complexes. The energy spectrum also matches qualitatively with recent single-molecule spectra of LH2 at low temperature. In this work we demonstrate that general features of organization are more important than the fine details of the LH apparatus, which is very promising in terms of designing artificial photosynthetic analogs.

Appendix.

Construction of K:

$$K_{nm,jm} = -iV_{n,j} \tag{A1}$$

$$K_{nm,nj} = iV_{j,m} \tag{A2}$$

$$K_{nm,\mathrm{nm}} = -\Gamma^*_{nm} + i\Delta_{nm} + \frac{1}{2}(k_t\delta_{n,tr} + k_t\delta_{m,tr}) \tag{A3}$$

where the indexes of K are constructed by the one-to-one mapping

$12, 13, \ldots, 1N, 21, 23, \ldots, 2N, \ldots, N1, N2, \ldots, (N-1)N \rightarrow 1,2, \ldots, N^2 - N$.

Construction of B:

Let B' be an (N - 1) x ($N^2$ – N) matrix, defined as follows

$$B' = \begin{bmatrix} V_{12} & \cdots & V_{1,N^2-N} \\ \vdots & \ddots & \vdots \\ V_{12} & \cdots & V_{1,N^2-N} \end{bmatrix} \tag{A4}$$

B is constructed by retaining only the elements, nm, of the $i^{th}$ row of B' in which either $n = i$ or $m = i$. In the former case, the sign of the nm element is '+', while in the latter it is '-'. All other elements are set to zero.

Expression for $\tau_{tr}$:

From (2), the expression for $\dot{\rho}_{tr}$ contains an additional term $k_t\rho_{tr}$. Therefore, after time integration and assuming that all the excitation eventually reaches the trap, we get

$1 = \mathrm{k_t\tau_{tr}}$.

Figures

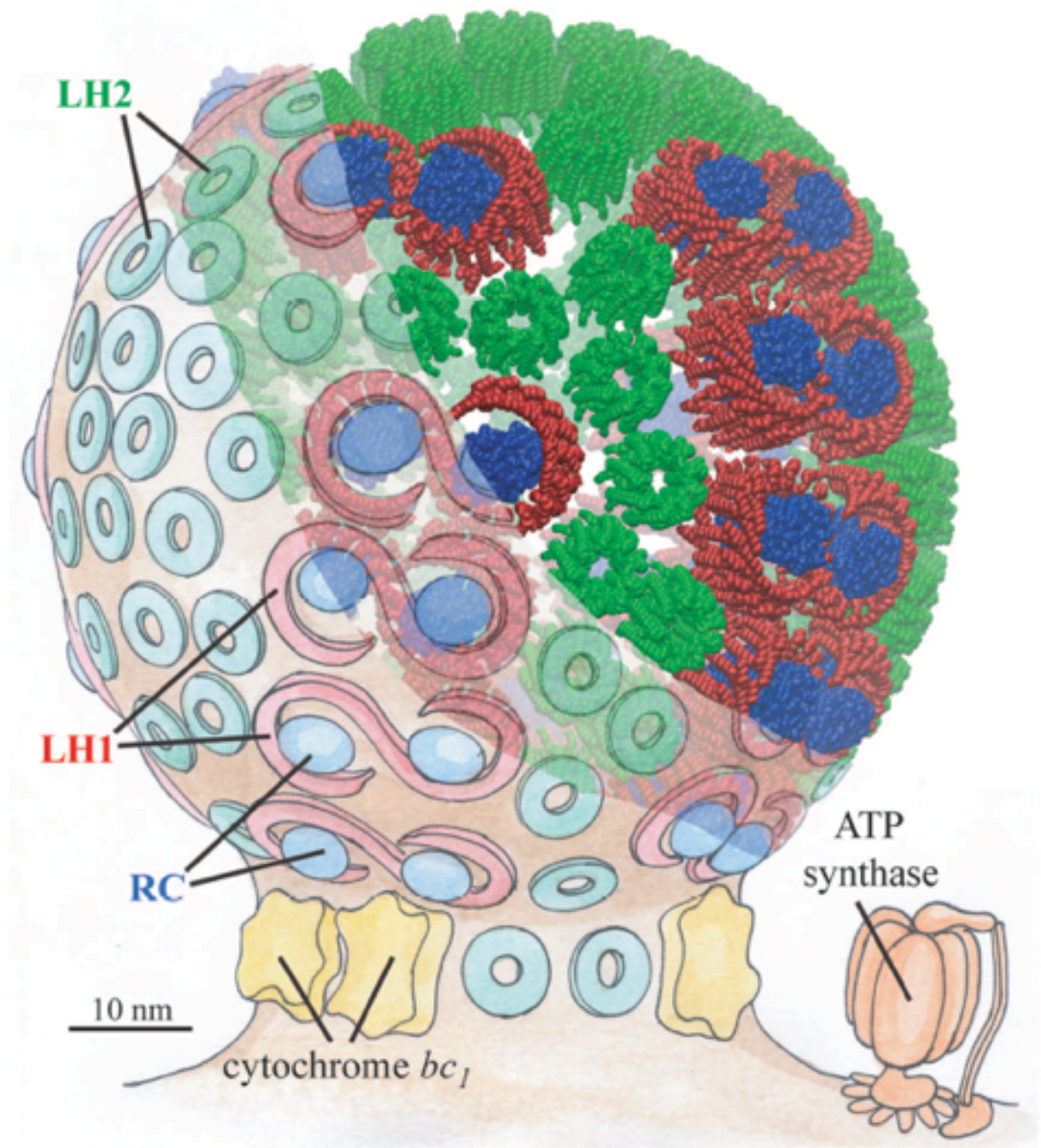

Figure 1: Light-harvesting apparatus in purple bacteria. Illustration of spherical chromatophore vesicle from *R. sphaeroides* showing organization of light harvesting complexes, LH2, and light-harvesting - reaction center complex (LH1-RC). Architecture and arrangement of constituent chromophores based on AFM images. Figure used with permission from Sener et al.[16]

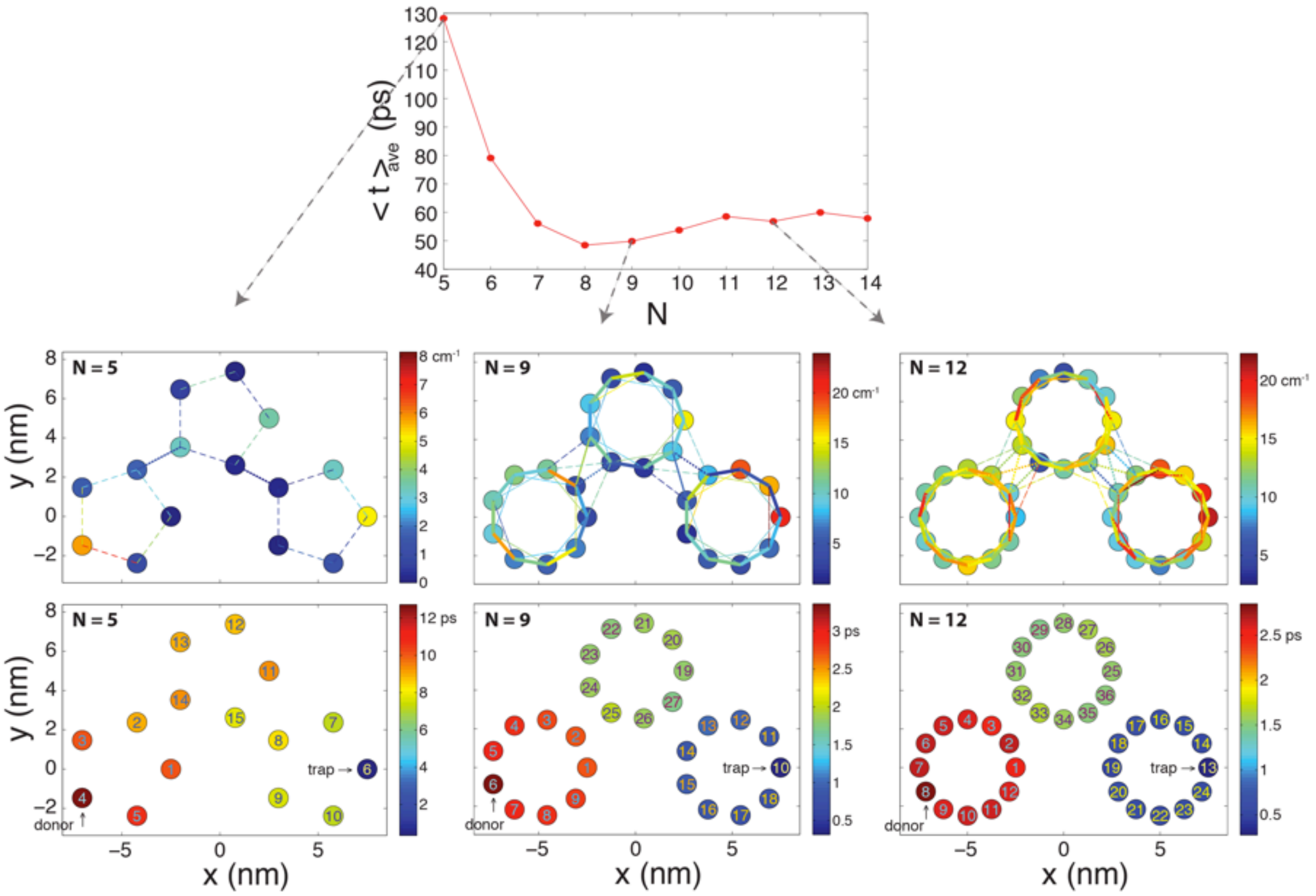

Figure 2. Optimal number of elements per ring as found by a genetic algorithm. Average trapping time as a function of the number of elements, N, in each ring. The diameter of each ring is 5 nm. The top row of images shows the strength of electrostatic coupling (lines between sites). Dashed lines correspond to > 5 $cm^{-1}$ and < 10 $cm^{-1}$, dotted lines to > 10 $cm^{-1}$ and < 20 $cm^{-1}$, and thick lines to > 20 $cm^{-1}$. Transition dipole at each site is normal to the plane. Color of connecting lines indicates the dephasing rate between two sites. Color of circles at each site corresponds to their energies. The bottom row of images shows the average residence time at each site, indicated by the color of the circles. The average trapping time is the sum of the average residence times at each site. The donor and trap states are labeled.

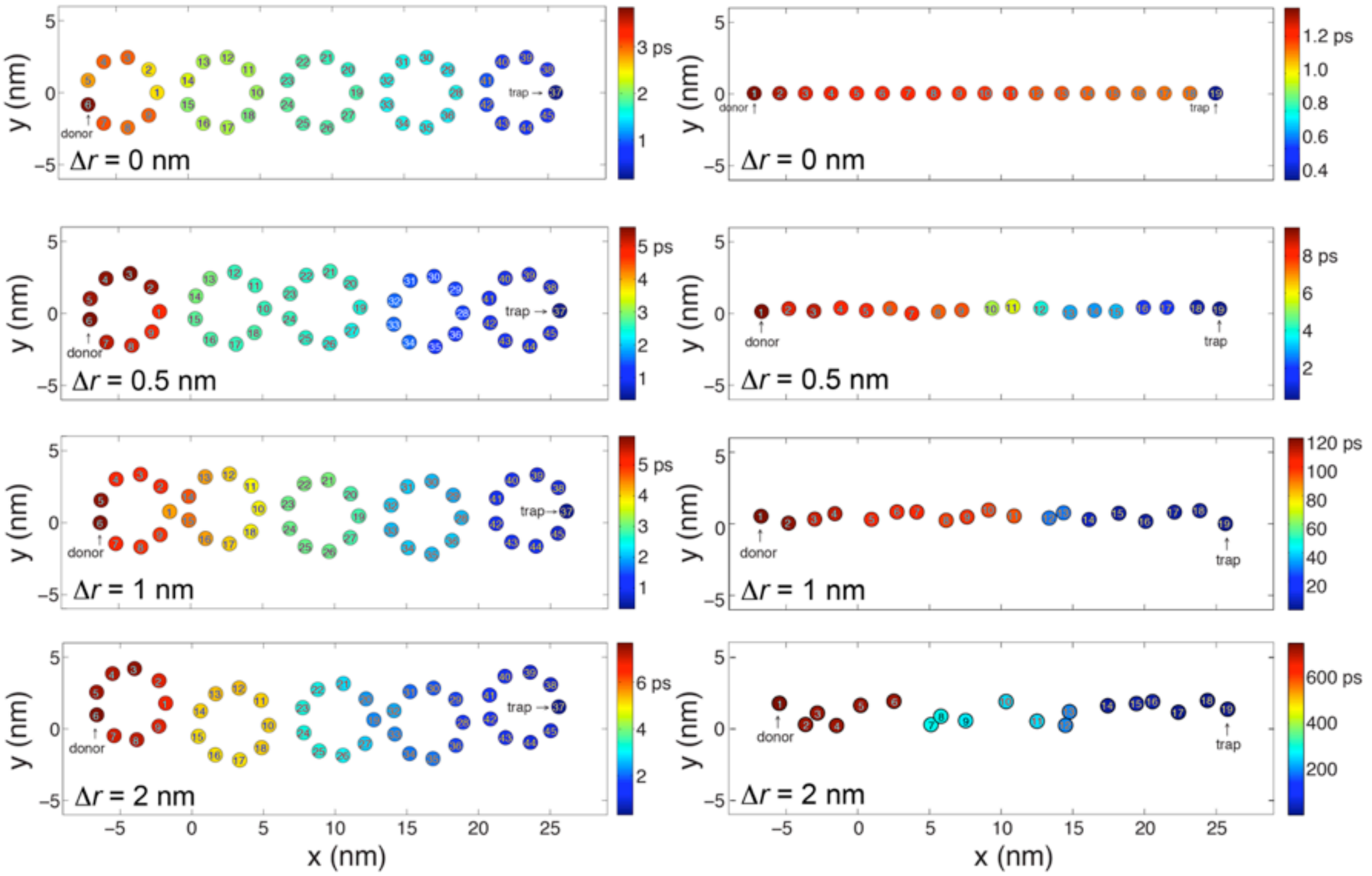

Figure 3: Chain of rings versus chain of individual chromophores. Comparison of the disorder between rings (left) and between chromophores (right) in a linear arrangement. Total distance from donor to trap is approximately the same in each case (~32 nm). Δr is the maximum, random displacement in both the x and y direction.

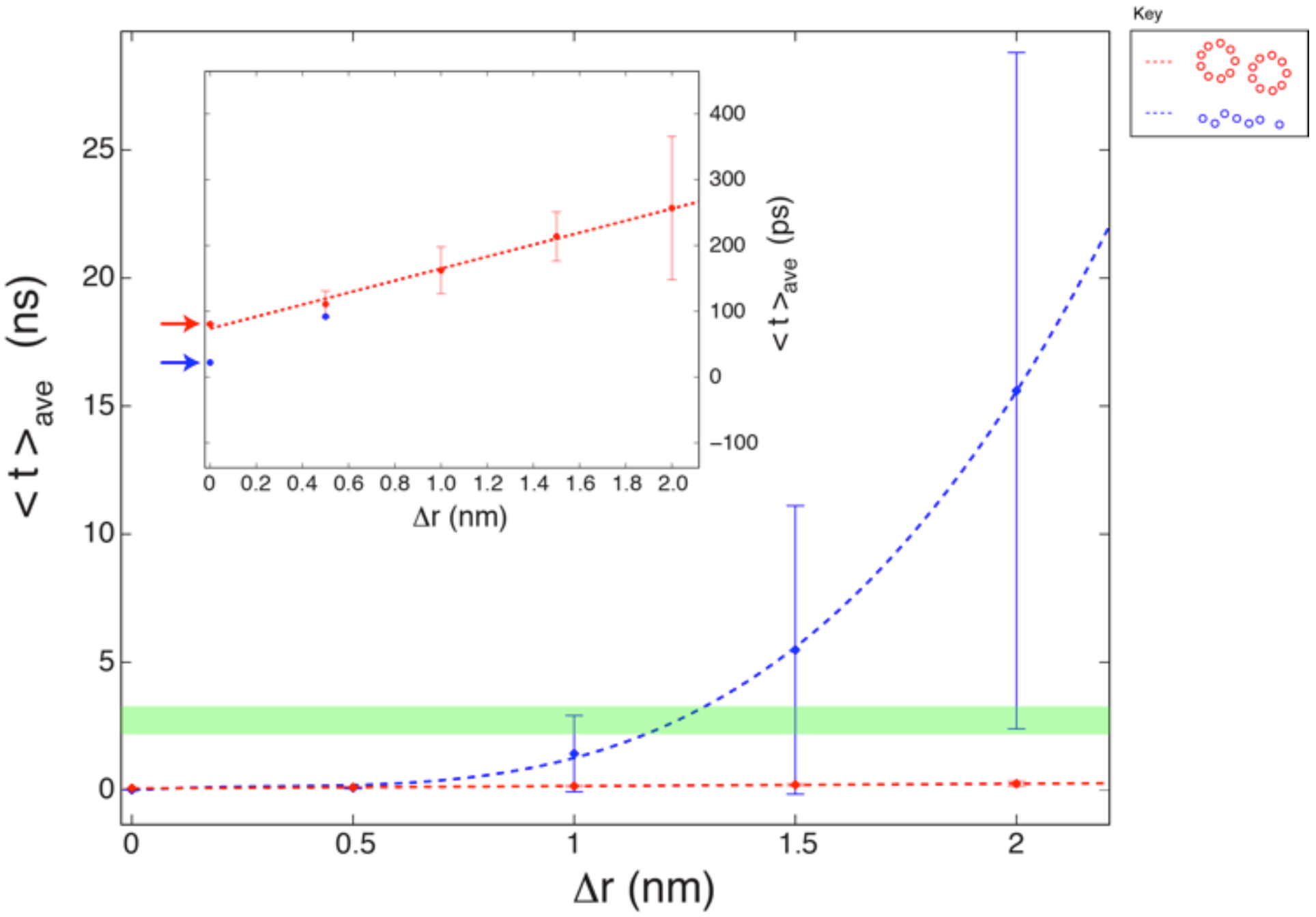

Figure 4: Rings are robust to spatial disorder. Plot of average trapping time versus static disorder between rings (red) and single sites (blue) from figure 3. Error bars for ring arrays are based on five runs through the optimization code. Error bars for array of single sites based on twenty runs through the optimization code. Outliers larger than $5\sigma$ were removed from the analysis. Single site arrays were fit to third-order polynomial, while ring arrays were fit to a least squares regression line. Inset shows a narrower window of trapping times (maximum of 400 ps). Arrows on the left of the inset indicate the mean residence time with no spatial disorder between rings. The green bar indicates approximate excited-state lifetime of Bacteriochlrophyll a, which represents the upper limit of relaxation of the sites back to the ground state. The linear array of sites performs better than the linear array of rings with no spatial disorder, but is significantly less robust to imperfections in positioning.

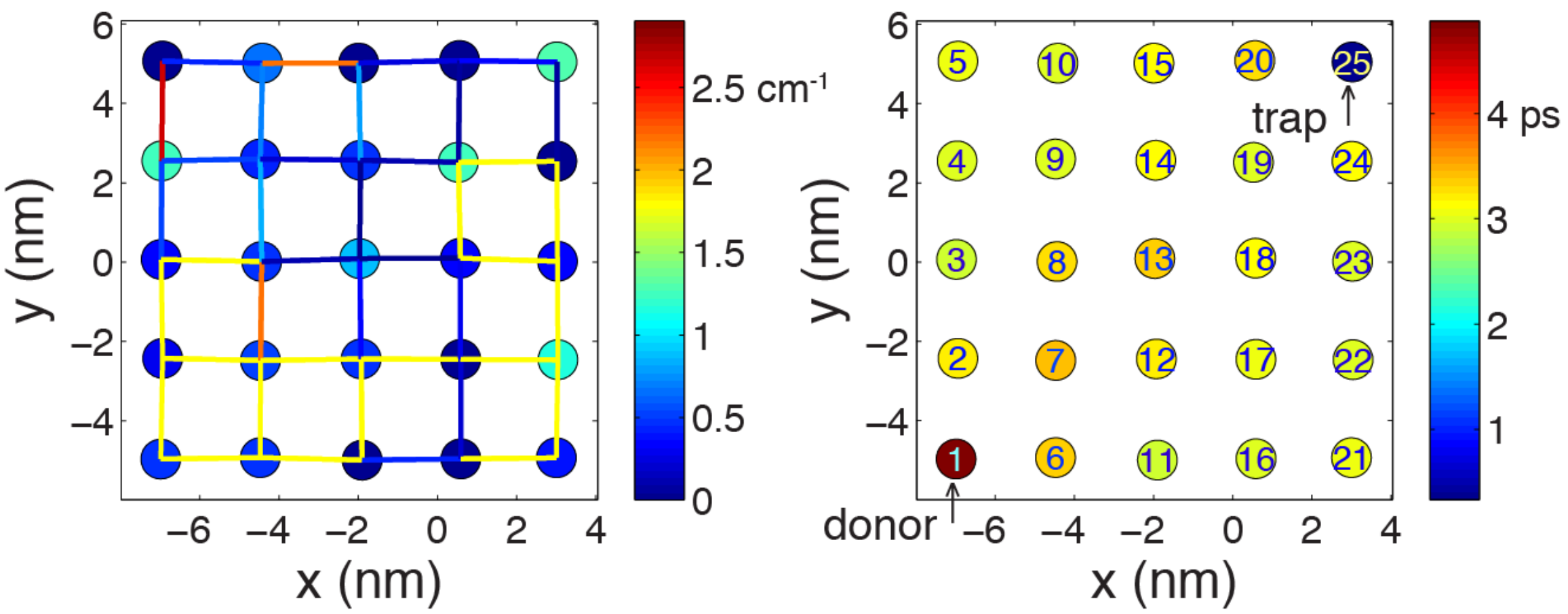

Figure 5: Grid-like array of chromophores is highly inefficient. Excitonic transfer through a grid-like arrangement of sites after optimization. Left: Colors of circles represent site energies. Color of lines represent dephasing rate. Right: Color represent mean residence time at each site. A significant amount of time is "wasted" at sites that do not directly link the donor to acceptor, i.e. sites 7, 13, and 19.

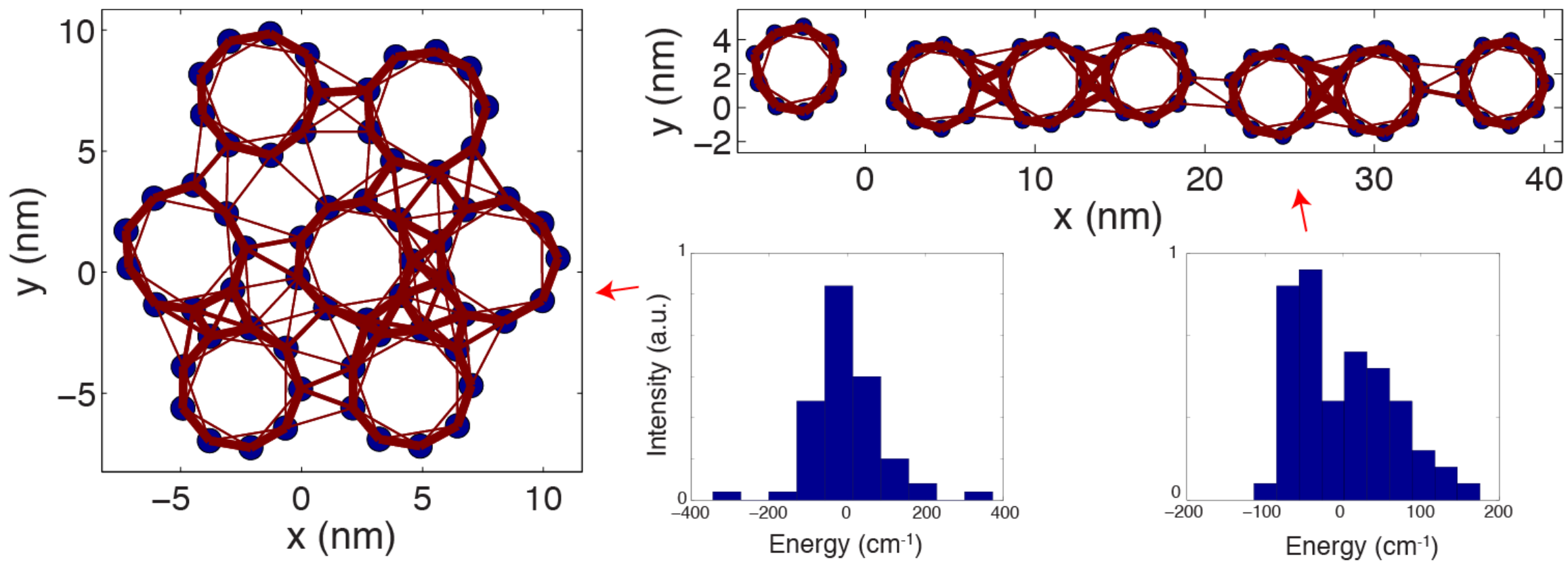

Figure 6: Two-dimensional packing is more robust to spatial disorder than linear chains. Electrostatic coupling between staggered (left) versus linear (right) in the presence of spatial disorder between rings for a 7-ring system. Thin lines indicate weak coupling (> 5 $cm^{-1}$ and < 10 $cm^{-1}$), medium lines indicate intermediate coupling (> 10$cm^{-1}$ and < 20 $cm^{-1}$) and thick lines indicates strong coupling (> 20 $cm^{-1}$). Staggered arrangement maintains non-negligible coupling strength and hence a path from donor to acceptor through multiple, neighboring rings. Linear arrangement more easily forms breaks, which may effectively block energy transfer across large distances. Bottom right: spectrum calculated by diagonalizing the system Hamiltonian. Energy spans approximately 200 $cm^{-1}$ in each case – a major role of coupling is to break site degeneracy and broaden the spectrum for efficient absorption of solar flux.

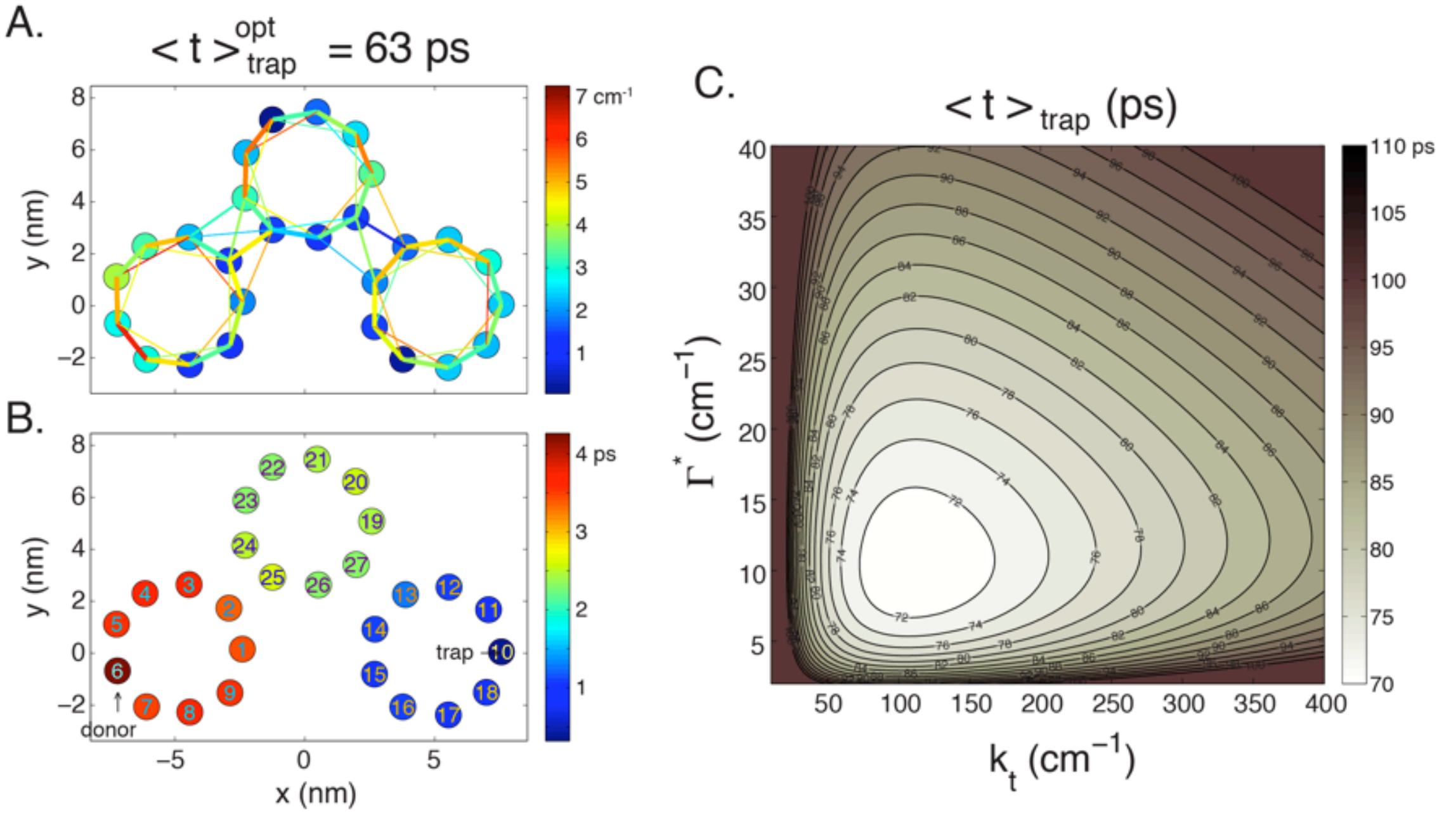

Figure 7: Near-optimal tapping is achieved without fine-tuning the system and bath. Left: Exciton transfer optimization achieved by a genetic algorithm to minimize the average trapping time as a function of the mutual dephasing between sites, trapping rate, and site energies. In the case of three rings in this arrangement, the optimal trapping time was found to be 63 ps. Right: Keeping the dephasing between sites constant and the site energies identical, the optimal trapping time is found to be ~70 ps. This indicates that fine-tuning of the system and the system-bath interactions is not necessary to achieve near-optimal transfer efficiency.

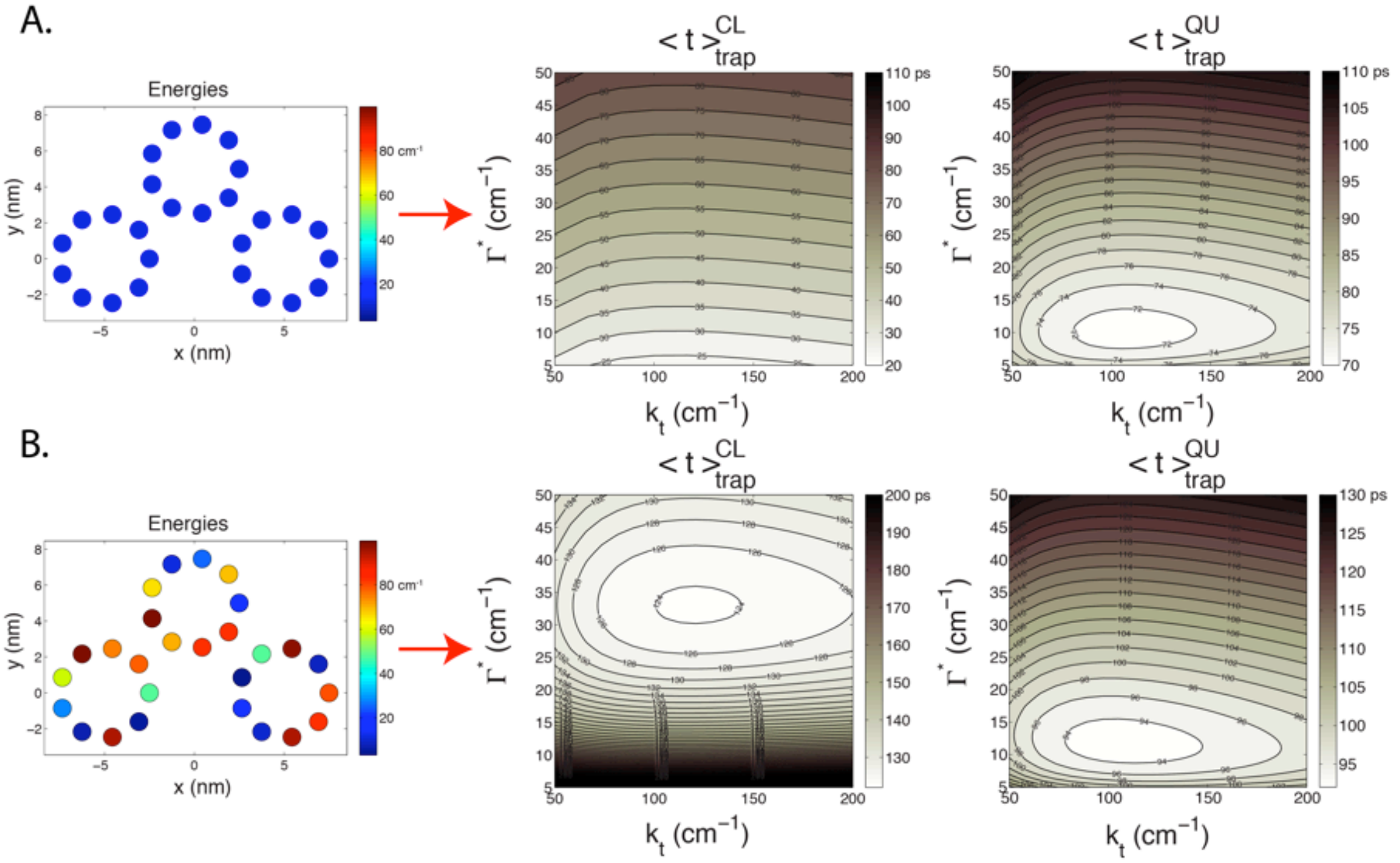

Figure 8: Quantum transport is robust to static energetic disorder. Comparison of classical and quantum transport in the absence (A) and presence (B) of static energetic disorder – difference in energies at each site. A. When the site energies are degenerate, classical transport predicts a shorter trapping time – near zero line width. Quantum transport predicts a slower trapping time by about a factor of three, but at a modest value of the dephasing rate. B. When the site energies are non-degenerate, classical transport undergoes a dramatic shift in the optimal line width. For the quantum case, changes in the optimal dephasing rate and trapping rate are negligible. In this case, the quantum transport is faster and significantly more robust to changes in site energies.